\documentclass[aps,prb,showpacs,twocolumn,reprint,superscriptaddress]{revtex4-1}
\usepackage{amsmath}
\usepackage{amssymb}
\usepackage{graphicx}
\usepackage{hyperref}
\begin{document}
\title{Multiferroic Properties of CaMn$_7$O$_{12}$}
\author{Guoquan Zhang}
\affiliation{Laboratory of Solid State Microstructures, Nanjing University, Nanjing 210093, China}
\author{Shuai Dong}
\email{sdong@seu.edu.cn}
\affiliation{Department of Physics, Southeast University, Nanjing 211189, China}
\author{Zhibo Yan}
\affiliation{Laboratory of Solid State Microstructures, Nanjing University, Nanjing 210093, China}
\author{Yanyan Guo}
\affiliation{Laboratory of Solid State Microstructures, Nanjing University, Nanjing 210093, China}
\author{Qinfang Zhang}
\affiliation{Computational Condensed Matter Physics Laboratory, RIKEN, Wako, Saitama 351-0198, Japan}
\affiliation{CREST, Japan Science and Technology Agency (JST), Kawaguchi, Saitama 332-0012, Japan}
\author{Seiji Yunoki}
\affiliation{Computational Condensed Matter Physics Laboratory, RIKEN, Wako, Saitama 351-0198, Japan}
\affiliation{CREST, Japan Science and Technology Agency (JST), Kawaguchi, Saitama 332-0012, Japan}
\author{Elbio Dagotto}
\affiliation{Department of Physics and Astronomy, University of Tennessee, Knoxville, Tennessee 37996, USA}
\affiliation{Materials Science and Technology Division, Oak Ridge National Laboratory, Oak Ridge, Tennessee 32831, USA}
\author{J.-M. Liu}
\email{liujm@nju.edu.cn}
\affiliation{Laboratory of Solid State Microstructures, Nanjing University, Nanjing 210093, China}
\affiliation{International Center for Materials Physics, Chinese Academy of Sciences, Shenyang 110016, China}
\date{\today}

\begin{abstract}
We report that CaMn$_7$O$_{12}$ is a new magnetic multiferroic material.
The appearance of a ferroelectric polarization coinciding with the magnetic phase transition ($\sim90$ K) suggests the presence of ferroelectricity induced by magnetism, further confirmed by its strong magnetoelectric response. With respect to other known magnetic multiferroics, CaMn$_7$O$_{12}$ displays attractive multiferroic properties, such as a high ferroelectric critical temperature and large polarization. More importantly, these results open a new avenue to search for magnetic multiferroics in the catalogue of doped oxides.
\end{abstract}
\pacs{75.80.+q, 75.47.Lx, 75.30.Kz}
\maketitle

\section{Introduction}
Multiferroics, with coexisting ferroelectric (FE) and magnetic orders
that are mutually coupled, have attracted considerable interest for
their technological relevance and fundamental science challenges.\cite{Cheong,Fiebig,Wang}
To avoid the natural exclusion between ferroelectricity and magnetism,\cite{Hill}
there are several routes to achieve multiferroicity.
Based on the microscopic origin of the FE polarization ($P$),
multiferroics can be classified into two families.\cite{Khomskii}
Type-I multiferroics, where ferroelectricity and magnetism have different origins,
often present high critical temperatures ($T_{\rm C}$'s) and polarizations.
However, the coupling between magnetism and ferroelectricity is usually weak.
In contrast, type-II multiferroics (i.e. magnetic multiferroics),
where the ferroelectricity is caused by a particular magnetic order,
are more interesting and important since both orders tend to be strongly coupled.

Since the discovery of TbMnO$_3$,\cite{Kimura} several magnetic multiferroics have been found.
However, in the spiral spin multiferroics,
the FE polarizations are regulated by the Dzyaloshinskii-Moriya (DM) interaction,\cite{Sergienko1,Katsura}
which originates from the spin-orbit coupling and, thus, is very weak.
This problem can be partially overcome in exchange-striction multiferroics.
For instance, the predicted FE $P$ in the E-type AFM manganites is of the order
of $10000$ $\mu$C/m$^2$,\cite{Sergienko2,Picozzi}
which has been recently confirmed in experiments.\cite{Ishiwata,Pomjakushin,Han}
However, the FE $T_{\rm C}$'s of the E-type AFM manganites remain low (typically $\sim30$ K)\cite{Ishiwata}
due to the magnetic frustration, namely the competition between nearest-neighbor (NN)
and next-nearest-neighbor (NNN) exchange interactions.\cite{Dong-frus,Mochizuki}

Among all the magnetic multiferroics, the narrow-bandwidth perovksite manganites define a very fertile field.
In recent years, the undoped $RE$MnO$_3$ case ($RE$ = Rare Earth) has been intensively studied,
both theoretically and experimentally. In addition,
recent theoretical calculations predicted that {\it doped} manganites may provide additional
magnetic multiferroic phases with even better physical properties.\cite{Giovannetti,Dong-Prl}
The expected spin structures of the new multiferroics are rather complex, involving zigzag-chains,
and they are stabilized by mechanisms that do not involve superexchange frustration,
such as electronic self-organization into stripes\cite{Dong-Prl} or correlation effects,\cite{Giovannetti}
that may be strong enough to obtain a high FE $T_{\rm C}$.
Also, the origin of the FE $P$ in these systems involves several aspects:
not only a DM contribution from noncollinear spin pairs can be present,
but also additional components triggered by spin dimmers and charge-orbital
ordering are available.\cite{Giovannetti,Dong-Prl} Clearly,
it would be important to test the general prediction of multiferroic complex states in real doped manganites.

In this manuscript, we will study the quadruple manganite CaMn$_7$O$_{12}$,
which is an ideal candidate to host new magnetic multiferroic phases.
From the structural point of view, CaMn$_7$O$_{12}$ belongs to the quadruple (AA'$_3$)B$_4$O$_{12}$ family,
which becomes more explicit by writing the chemical formula as
(Ca$^{2+}$Mn$^{3+}_3$)(Mn$^{3.25+}_4$)O$^{2-}_{12}$.\cite{Vasilev}
This perovskite-derived structure consists of a three-dimensional array of
corner-sharing BO$_6$ octahedra, which are considerably tilted due to the
small size of the A-site (Ca$^{2+}$Mn$^{3+}_3$), as shown in Fig.~1(a).
Therefore, the B-site Mn-O-Mn bond is short and bended, giving rise to a
very narrow bandwidth system with robust DM interaction, considered
very important in multiferroic manganites. In addition, different from
normal perovskite manganites, the Jahn-Teller (JT) distortion ($Q_2$ mode)
in CaMn$_7$O$_{12}$ is weak.\cite{Przenioslo:Pb} This weak $Q_2$ mode
and a moderate Jahn-Teller $Q_3$ mode can stabilize some multiferroic phases
(such as the predicted ``SOS'' phase)\cite{Dong-Prl} which are difficult
to form in normal narrow bandwidth perovskite manganites with strong JT distortion.
Furthermore, in this quadruple structure, the A-site Ca$^{2+}$Mn$^{3+}_3$ are fully
ordered, different from the random distribution of A-site cations in standard chemically doped
perovskites. The reduction of quenched disorder provides an extra advantage to stabilize
complex spin patterns at a relatively high temperature ($T$).

Quadruple perovskite manganites have not been much
studied,\cite{Vasilev,Prodi1,Prodi2} particularly with regards to their
multiferroicity. Imamura \textit{et al}. and Mezzadri \textit{et al}.
reported  room-$T$ ferroelectricity (evidenced from the cation displacements)
in BiMn$_7$O$_{12}$.\cite{Imamura,Mezzadri}
However, its polarization is difficult to measure due to its low resistivity at room-$T$.
Since the ferroelectricity in BiMn$_7$O$_{12}$ is caused by the well-known $6s^2$
lone pair of Bi$^{3+}$ instead of a magnetic order, then this material
is not a magnetic multiferroic. As for CaMn$_7$O$_{12}$,
S\'anchez-And\'ujar \textit{et al}. reported a magnetodielectric effect at low $T$.\cite{Andujar}
However, it has been well recognized that a magnetodielectric effect may be unrelated to a true magnetoelectric coupling.\cite{Catalan:Apl}

\begin{figure}
\includegraphics[width=0.4\textwidth]{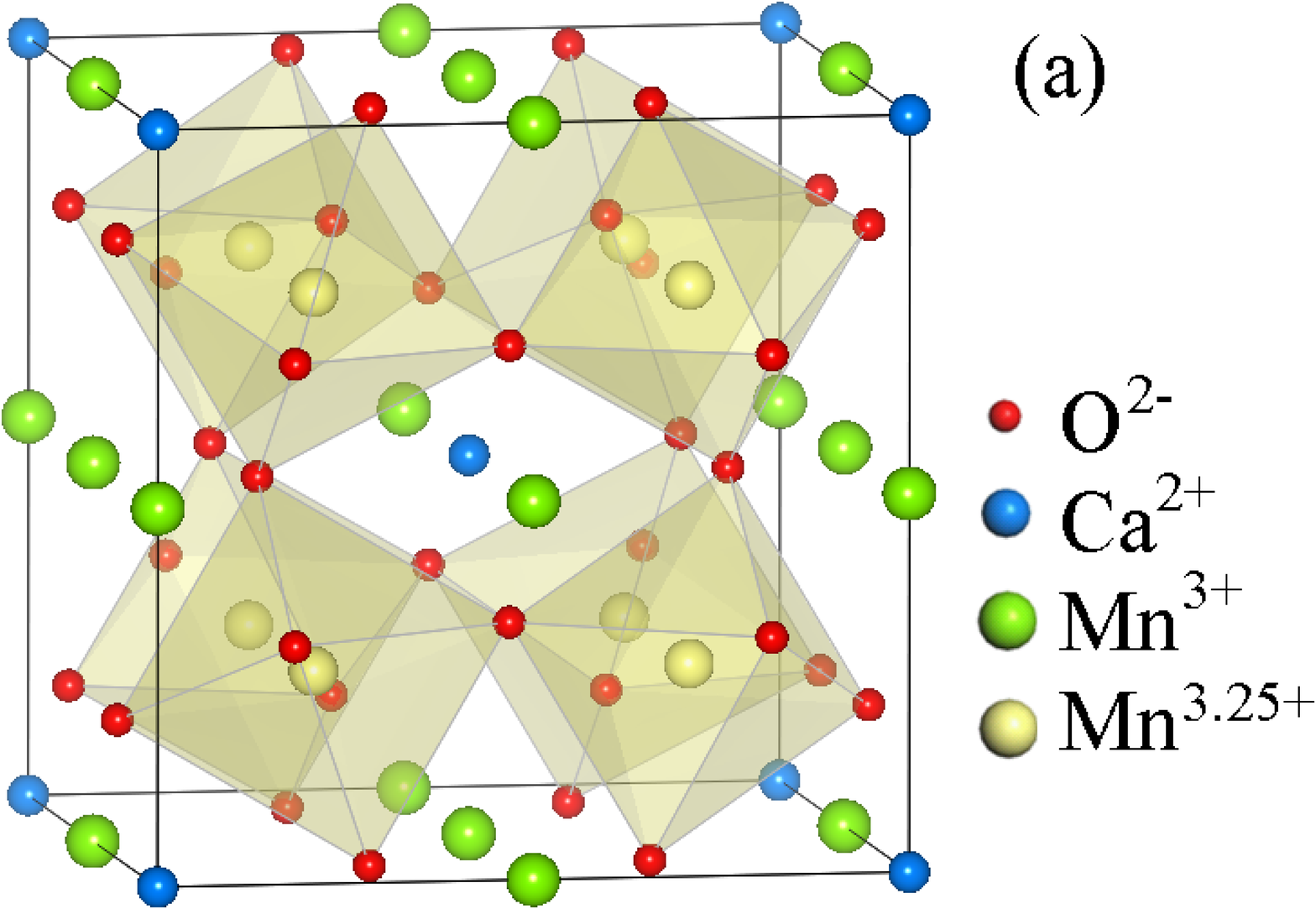}
\vskip -0.6cm
\includegraphics[width=0.45\textwidth]{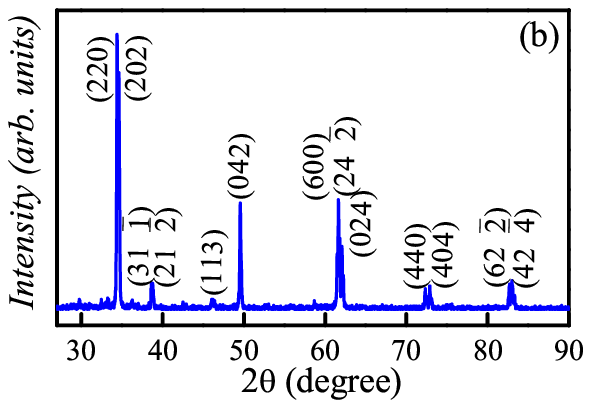}
\vskip -0.6cm
\caption{(color online) (a) Sketch of the crystal structure of CaMn$_7$O$_{12}$.
The A(A')-site Ca$^{2+}$/Mn$^{3+}$ cations are ideally ordered.
The lattice is cubic (Im\={3} with a lattice constant $a\approx7.35$ \AA) above $\sim440$ K, while
below this temperature it distorts into a rhombohedral arrangement
(R\={3} with lattice constants $a\approx10.44$ \AA{} and
$c\approx6.34$ \AA).\cite{Vasilev,Przenioslo:Pb}
The Mn$^{3.25+}$ cations locate in the center of oxygen octahedra,
which become (nominal) Mn$^{3+}$/Mn$^{4+}$ charge ordered below $250$ K.\cite{Przenioslo:Pb}
Due to the small size of Ca$^{2+}$Mn$^{3+}_3$, the lattice is shrunk and the oxygen octahedra
are more tilted when compared with other perovskite manganites.
For example, the NN distance of B-site Mn's
(nominal bond length) is only $\sim3.68$ \AA{} and the B-site Mn-O-Mn bond angle
is $\sim138^\circ$,\cite{Przenioslo:Pb} both much smaller than those of standard
perovksite manganites (e.g. $\sim3.88$ \AA{} and $\sim160^\circ$
for La$_{0.75}$Ca$_{0.25}$MnO$_3$).\cite{Radaelli}
(b) The room-$T$ XRD spectrum.}
\vskip -0.6cm
\end{figure}

\section{Methods}
Polycrystalline samples of CaMn$_7$O$_{12}$ were prepared using the standard sol-gel method.
Ca(NO$_3$)$_2\cdot$4H$_2$O and Mn(NO$_3$)$_2$ ($50\%$) were chosen as reagents and treated in the
same way as reported in previous literature.\cite{Shankar:Ssc,Li:Mseb} To avoid the impurities
Mn$_2$O$_3$, Mn$_3$O$_4$ or maybe CaMnO$_3$, the obtained sol-gel precursors were ground
and then heated in air at $800$ $^\circ$C/$48$ h, $925$ $^\circ$C/$48$ h, and $950$ $^\circ$C/$48$ h,
respectively, with intermediate grindings. The resultant powders were pressed into pellets
and sintered in air at $970$ $^\circ$C for $60$ h.
As a result of a careful preparation, the samples are found to be in a single phase, as shown in Fig.~1(b). The average grain size (diameter) is $\sim1.0$ $\mu$m according to the scanning electron microscope (SEM) micrograph (not shown).

The sample crystallinity was characterized by X-ray diffraction (XRD) with Cu $K\alpha$ radiation
at room temperature. The dc magnetic susceptibility ($\chi$) as a function of $T$ was measured
using the Quantum Design superconducting quantum interference device (SQUID), and specific-heat
measurements using the ``physical properties measurement system'' (PPMS) were performed.
To measure the dielectric constant $\varepsilon$ and polarization $P$, gold pastes were used as
electrodes while the varying $T$ and magnetic field $H$ environment was provided by PPMS.
The $\varepsilon-T$ data were collected using the HP4294 impedance analyzer at various frequencies.
The polarization $P$ as a function of $T$ was measured using the pyroelectric current method plus
a careful exclusion of other possible contributions, such as those from the de-trapped charges.
The samples were first poled under selected electric fields from $120$ K to $8$ K, and then the
pyroelectric currents, using the Keithley 6514A electrometer, were integrated by warming
the sample at different ramping rates of $2\sim6$ K/min, respectively. At present, this
pyroelectric currents process is the most widely used method to measure ferroelectricity
in magnetic multiferroics, since their polarizations are usually too weak
(typically $3-4$ orders of magnitude weaker than traditional ferroelectrics)
to be verified by FE hysteresis loops.

\section{Results and Discussion}
Let us first consider the magnetic properties. The $T$-dependence of the
magnetic susceptibility ($\chi$) is shown in Fig.~2(a). Upon cooling, first,
a small kink in $\chi$ appears at $T_{\rm N1}\sim90$ K, indicating an AFM transition,
in agreement with previous literature.\cite{Przenioslo:Pb,Przenioslo:Ssc,Przenioslo:Apa,Presniakov}
When the sample is further cooled down to $T_{\rm N2}\sim48$ K, $\chi$ rapidly grows.
Below $T_{\rm N2}$, the zero-field-cooling (ZFC) and field-cooling (FC) $\chi$'s diverge
from each other. Using the Curie-Weiss law to fit the data above $\sim90$ K,
a good paramagnetic (PM) behavior above $T_{\rm N1}$ is observed, as indicated by the
dashed line in Fig.~2(b). The extrapolated
Curie-Weiss temperature $\Theta$ is $\sim$$-50$ K.
Such a Curie-Weiss behavior and a negative $\Theta$ confirm the PM-AFM transition
at $T_{\rm N1}$. The phase transitions at $T_{\rm N1}$ and $T_{\rm N2}$ are further
confirmed by the specific heat $C_{\rm p}$, which exhibits sharp anomalies at both $T$'s, as shown in Fig.~2(a).
In addition, magnetic hysteresis loops were measured, as shown in Fig.~2(c),
indicating a weak ferromagnetic (FM) signal at low $T$'s.

\begin{figure}
\vskip -1cm
\includegraphics[width=0.5\textwidth]{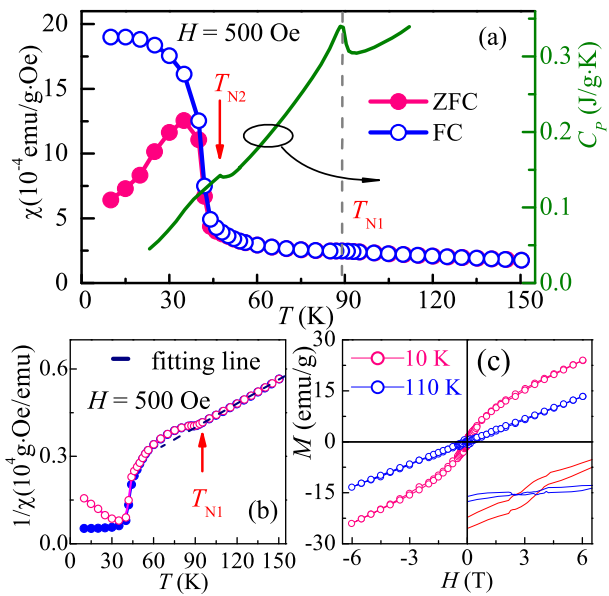}
\vskip -0.6cm
\caption{(color online) (a) The magnetic susceptibility ($\chi$, left axis,
FC and ZFC) and specific heat ($C_{\rm p}$, right axis) versus $T$.
(b) Inverse of $\chi$ (FC and ZFC) versus $T$.
The Curie-Weiss law provides a good fit above $T_{\rm N1}$.
(c) The magnetic hysteresis loops at two typical $T$'s.
The loops show PM behaviors (linear $M-H$ relationship without a coercive field)
above $T_{\rm N1}$ and a weak FM-like signal (with a coercive field $\sim450$ Oe)
at a low $T$. The inset in the fourth quadrant is a zoomed view near $H=0$.
The horizontal and vertical scales are [$-3000$, $3000$] Oe and [$-5$, $5$] emu/g, respectively.}
\vskip -0.6cm
\end{figure}

\begin{figure}
\vskip -0.8cm
\includegraphics[width=0.5\textwidth]{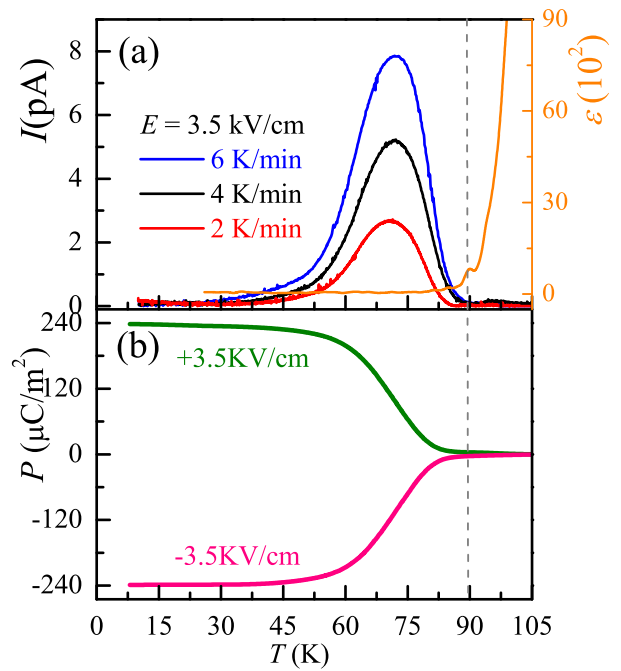}
\vskip -0.6cm
\caption{(color online) (a) The pyroelectric currents, under different warming rates,
($I$, left axis) and the dielectric constant ($\varepsilon$, right axis) vs. $T$.
The poling electric field is $3.5$ kV/cm. Note that there is a peak in the
$\varepsilon-T$ curve at $T_{\rm N1}$. (b) The (symmetric) FE polarizations under
positive and negative poling electric fields.}
\vskip -0.6cm
\end{figure}

The dielectric and FE properties were the main focus of this effort.
First, the dielectric constant $\varepsilon$ as a function of $T$,
was measured at frequency $f=1$ kHz, as shown in Fig.~3(a). A small but clear
anomaly in $\varepsilon(T)$ was identified at $90$ K indicating a FE transition,
coinciding with $T_{\rm N1}$, the AFM transition point.
Even more interesting, as shown below a FE polarization emerges at this AFM transition,
suggesting the presence of ferroelectricity induced by magnetism.
In other words, CaMn$_7$O$_{12}$ is here shown to be a magnetic multiferroic material.
The reliability of the measured $P(T)$ is evidenced by
the pyroelectric current ($I$) as a function of $T$
with different warming rates: $2$ K/min, $4$ K/min, and $6$ K/min, as indicated
in Fig.~3(a). The peaks of the three $I-T$ curves do not shift along the $T$-axis,
and the integral of $I(t)$ for the three curves are almost identical (not shown),
indicating that the de-trapped charges, if they exist, do not contribute appreciably
to the measured current and, then, the intrinsic pyroelectric current is dominant.
We also measured $P(T)$ under positive and negative poling electric fields $E=\pm3.5$ kV/cm,
respectively. The polarity of $P$ is determined by the sign of $E$,
as shown in Fig.~3(c), confirming the existence of ferroelectricity in CaMn$_7$O$_{12}$.

Under the poling field $E=3.5$ kV/cm, the measured $P$ reaches up to $\sim240$ $\mu$C/m$^2$ at $8$ K. However, the real saturated FE $P$ is much larger due to the high coercive field. For example, by using a larger poling field $E=7$ kV/cm (the highest field we can apply in the current stage),
the FE $P$ increases to $\sim450$ $\mu$C/m$^2$ at $8$ K.
The $E$-dependence of $P$ at $8$ K is presented in Fig.~4(b). Since the $P-E$ curve
does not saturate up to $E=7$ kV/cm, a larger saturated $P$ is to be expected, which
could be measured with higher poling fields (or using thin films).\cite{Pomjakushin,Han}
In addition, a tiny anomaly in the pyroelectric current emerges at $T_{\rm N2}$ under $E=7$ kV/cm,
as shown in Fig.~4(a), which gives rise to a small increase of $P$.
However, this anomaly at $T_{\rm N2}$ is not distinct under $E=3.5$ kV/cm, as shown in Fig.~3(a).

The existence of magnetic multiferroicity has been further confirmed
by the presence of a strong magnetoelectric (ME) response, as shown by the magnetic field,
$H$, dependence of $P$. During the measurement, the sample is cooled from $120$ K to $8$ K
under various $H$'s and using a fixed $E=3.5$ kV/cm. As shown in Fig.~4(c),
the measured $P$ is suppressed by the applied $H$, e.g. it decreases to $\sim160$ $\mu$C/m$^2$
at $8$ K under $H=9$ T, while $T_{\rm c}$ is almost unchanged which suggests a robust magnetic transition.
The ME response, defined as $(P(0)-P(H))/P(0)\times100\%$, reaches up to $30\%$ under $H=9$ T
(see inset of Fig.~4(c)). Considering the polycrystalline nature of the sample,
this ME response is large (i.e. comparable with those of $RE$MnO$_3$
($RE$=Eu$_{1-x}$Y$_{x}$ and Lu) polycrystals),\cite{Ishiwata}
implying the presence of magnetism-induced ferroelectricity.

\begin{figure}
\vskip -0.4cm
\includegraphics[width=0.5\textwidth]{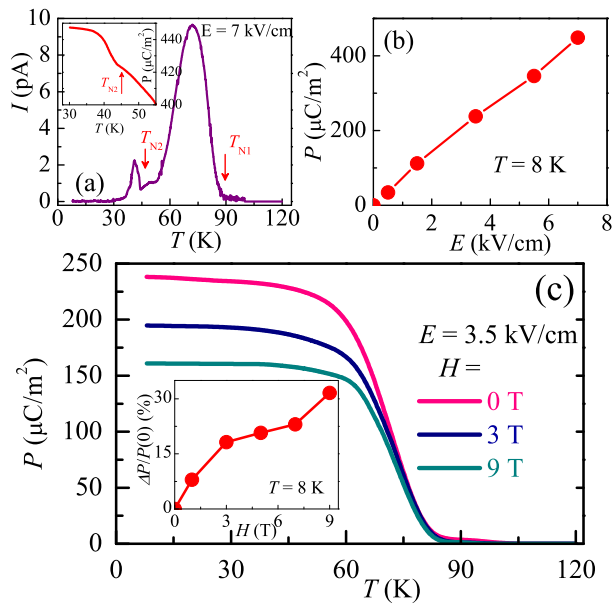}
\vskip -0.6cm
\caption{(color online) (a) The pyroelectric currents under a large $E$.
An anomalous contribution appears around $T_{\rm N2}$. Inset: The corresponding FE $P$
around $T_{\rm N2}$. (b) The FE $P$ at $8$ K vs. $E$, which does not saturate up to
$7$ kV/cm (the maximum field we can apply to the bulk sample).
(c) The suppression of FE $P$ by a magnetic field. Inset: the magnetoelectric response ratio.}
\vskip -0.4cm
\end{figure}

Compared with other magnetic multiferroics, the multiferroicity
of CaMn$_7$O$_{12}$ is remarkable. First, the observed FE $T_{\rm c}$ is considerably
higher than in typical magnetic multiferroics. For example, the $T_{\rm c}$'s
of orthorhombic $RE$MnO$_3$ ($RE$=Tb, Dy, Eu$_{1-x}$Y$_x$, Ho, Y, Tm, and Lu)
are all below $35$ K.\cite{Ishiwata}
Second, the observed $P$ (saturated value $>450~\mu$C/m$^2$) is quite large as compared with other magnetic multiferroics,
considering the polycrystalline nature of the sample. For $RE$MnO$_3$,
the FE $P$ of a high-quality polycrystal is estimated to be only $1/6$ of
the single crystal one.\cite{Ishiwata}
Therefore, the expected saturated $P$ of CaMn$_7$O$_{12}$ single crystal can be larger than $2700~\mu$C/m$^2$,
which is already larger than in DyMnO$_3$ ($\sim$$2000~\mu$C/m$^2$).
Even comparing with the E-AFM manganites (e.g. $\sim$$600-900~\mu$C/m$^2$ for high-quality polycrystals
with a poling field $E=8$ kV/cm),\cite{Ishiwata},
the value $450~\mu$C/m$^2$ (with $E=7$ kV/cm) of CaMn$_7$O$_{12}$ is remarkable.
The large $P$ of CaMn$_7$O$_{12}$ could originate from the following two mechanisms.
First, a robust DM interaction may exist in this material. As shown in Fig.~1(a),
the B-site Mn-O-Mn bonds are much more bending than those in normal perovskites,
which gives rise to a larger DM interaction.\cite{Sergienko1}
Second, the exchange striction\cite{Sergienko2,Picozzi}
may also exist in these materials.

Finally, it should be noted that the spin order
of CaMn$_7$O$_{12}$ is very complex.
Earlier neutron studies\cite{Przenioslo:Pb,Przenioslo:Ssc,Przenioslo:Jpcm}
found that the fittings using collinear spin modes
were not quite successful.\cite{Vasilev,Przenioslo:Pb}
Thus, it is crucial to carry out additional
high-precision neutron investigations
to figure out better the magnetic order of CaMn$_7$O$_{12}$,
which is essential to further clarify its microscopic multiferroic mechanism.

Recently, after our first submission of this manuscript, we noticed an experimental
investigation by Johnson \textit{et al.} %building on our finding of ferroelectricity in CaMn$_7$O$_{12}$
that also reported the existence of \emph{magnetic multiferroicity} in this compound,
and in addition they find a noncollinear spin order in this material via neutron studies.\cite{Johnson}
Their results agree with
our data quite well, as exemplified by the following facts:\\
(1) The FE $T_{\rm c}$'s are identical at $T_{\rm N1}$.\\
(2) The FE $P$ reported by Johnson {\it et al.} is quite large: up to $2870$ $\mu$C/m$^2$ for a
single crystal,\cite{note} which is about $6$ times our maximum value $450$ $\mu$C/m$^2$ for polycrystalline samples.\\
(3) The $P$-$T$ curves are also very similar.\\
(4) Their neutron studies suggest that the FE $P$ is induced by a noncollinear spin order, compatible with our report of a strong magnetoelectric response.\\
(5) A small anomaly in $P$ at $T_{\rm N2}$ was also noticed in their single crystal with a poling field $E=4.4$ kV/cm.\cite{note}\\
In Ref.~\onlinecite{Johnson}, the spin order between $T_{\rm N1}$ and $T_{\rm N2}$
has been resolved: this spin order is quite complex with a noncollinear structure
(involving both the A-site and B-site Mn cations) and it contradicts earlier neutron
diffraction results.\cite{Przenioslo:Ssc} However, the magnetic order below $T_{\rm N2}$ remains unclear.
Therefore, overall the magnetic multiferroic character of CaMn$_7$O$_{12}$ is nicely
confirmed by Johnson {\it et al.}, while the clarification of the
underlying microscopic mechanism still needs further experimental and theoretical careful investigations.

\section{Conclusion}
In summary, we have found that the quadruple perovskite CaMn$_7$O$_{12}$ is
a magnetic multiferroic material. Its multiferroic properties are interesting
(large-$P$, high-$T_{\rm c}$, and a strong magnetoelectric response) when compared with other known magnetic multiferroic manganites. A new physical mechanism
appears to be needed to explain our results.
Considered more broadly, these results for CaMn$_7$O$_{12}$ open a new route
to pursue higher-$T_{\rm c}$ and larger-$P$ magnetic multiferroics via the use of doped oxides.
This is interesting since most previous verified multiferroic materials are actually
undoped, since doping was expected to bring extra carriers and destroy the insulating behavior
required by ferroelectricity. Our results show that this is not necessarily
correct for narrow-bandwidth oxides.
Besides further investigating the properties of CaMn$_7$O$_{12}$, our effort suggests
that it would be important to search
for additional new magnetic multiferroics in the quadruple perovskite family.

\acknowledgments{We thank Silvia Picozzi for bringing quadruple manganites to our attention
and Riccardo Cabassi for helpful discussion. This work was supported by the 973 Projects
of China (2009CB623303, 2011CB922101) and the National Science Foundation of China
(50832002, 11004027). Q.F.Z and S.Y. were supported by CREST-JST. E.D. was supported by the
U.S. Department of Energy, Office of Basic Energy Sciences, Materials Sciences and Engineering Division.}


\begin{thebibliography}{a}
\bibitem{note} Due to the voltage-divider effect between grain boundaries in polycrystalline samples, the ``effective'' poling fields to FE domains are smaller than the nominal ones. Thus, the efficiency of a poling field is higher in single crystals than in polycrystals.
\end{thebibliography}
\end{document}